**Geographically Weighted Surrogate Models for Rapid Small-Area Chronic Disease Estimation**

Aanya Gupta[a], Szandra A. Péter[b], Sara Von Hoene[b], Emma Von Hoene[b], Taylor Anderson[b*]

[a]*Department of Biomedical Engineering, Duke University, Durham, NC, USA*

[b]*Department of Geography and Geoinformation Sciences, George Mason University, Fairfax, VA, USA*

***corresponding author**

Taylor Anderson tander6@gmu.edu

**Funding**

This project was funded by the National Science Foundation (Award No. 2109647).

**Data availability**

All data and code used in this study is available in a GitHub repository at https://github.com/AanyaAGupta/Geographically-Weighted-Surrogate-Models-for-Rapid-Small-Area-Estimation

**Geographically Weighted Surrogate Models for Rapid Small-Area Chronic Disease Estimation**

Anonymous

## Abstract

Small-area estimation (SAE) enables researchers and policymakers to identify spatial disparities in health outcomes, but survey-based SAE products carry an inherent lag. Gold-standard estimates such as CDC PLACES are released roughly two years after the underlying survey data are collected, limiting their use for time-sensitive decision-making. This study evaluates the potential for machine learning (ML) to serve as a surrogate, learning the relationship between frequently updated area-level predictors and existing SAE outputs to generate timely, comparable estimates in years when SAE from surveys are unavailable or delayed. We evaluate several global and geographically weighted ML models for county-level SAE of ten chronic conditions across the US: COPD, asthma, heart disease, arthritis, cancer, depression, diabetes, high blood pressure, high cholesterol, and stroke. Our findings suggest that geographically weighted ML frameworks like geographically weighted random forest and geographically weighted regression offer scalable and open data surrogates for rapidly generating SAE and supporting data driven decision making.

# Geographically Weighted Surrogate Models for Rapid Small-Area Chronic Disease Estimation

## 1. Introduction

Small-area estimation (SAE) refers to a collection of methods that combine survey and area-level data to generate reliable population estimates for small geographic areas, such as census tracts or counties (U.S. Census Bureau, 2026). In the United States, SAE plays a key role in public health surveillance by enabling researchers and policymakers to produce reliable estimates of health behaviors or outcomes for populations in regions with limited ground truth data. Accurate local estimates aid in identifying geographic health disparities, help target interventions, and inform evidence-based policy making (Pfeffermann, 2013; Rahman, 2017; Benavidez et al., 2024). Some states obtain direct SAE through county or regional-level representative surveys, but these efforts are not conducted uniformly across states and may not yield reliable direct estimates for every county. Furthermore, direct estimation is not feasible at granularities finer than county-level (e.g., tract, block group). Thus, model-based SAE is a standard way to fill these gaps in public health spatial data availability (Ghosh & Rao, 1994; Pfeffermann, 2013).

Model-based SAE encompasses a variety of methods. As summarized by Rahman (2017), SAE methods can be classified into several categories: multilevel statistical modeling, microsimulation, and machine learning-based approaches. Multilevel modeling fits a regression model using individual and area-level predictors to estimate probabilities of outcomes, which are then applied to population counts to generate area-specific estimates (CDC, 2024a; Umlauf et al., 2015; Zhang, 2013). Microsimulation modeling generates synthetic populations that preserve the joint distributions of sociodemographic variables which are then aggregated to estimate health outcomes at fine geographic scales (Von Hoene et al., 2025). More recently, machine learning (ML) approaches have been explored for SAE because of their ability to flexibly model non-linear relationships and interactions (Viljanen et al., 2022; Krennmair et al., 2022; Żądło & Chwila, 2025). In these frameworks, ML models are typically trained on individual- or household-level survey data to learn the relationship between covariates and outcomes, and predictions are then aggregated to the area level to produce SAEs.

A practical limitation shared by all survey-dependent SAE approaches, including the gold-standard multilevel regression and poststratification method used by CDC PLACES (CDC, 2024b) is the inherent lag between survey data collection and SAE estimate availability. For example, CDC PLACES estimates are released approximately two years after the underlying BRFSS (CDC, 2026) survey data are collected. This delay limits the utility of these products for time-sensitive public health decision-making, particularly in rapidly evolving public health crises. Importantly, publicly available area-level data, such as U.S. Census Bureau American Community Survey estimates, are updated annually and released with considerably less lag than survey-based SAE products. Prior work has also shown that these data can serve as reliable predictors in machine learning approaches for estimating small-area health outcomes (Feng & Jiao, 2021).

This suggests an opportunity for ML approaches to serve as surrogate models. In this study, we use the term *SAE surrogate model* to refer to a data-driven ML model trained to approximate existing complex survey-based SAE using more frequently updated area-level predictors. Rather than replacing survey-based SAE or estimating health outcomes directly from individual-level survey microdata, the proposed framework emulates the outputs of established SAE products, enabling timely and comparable estimates when official releases are delayed or unavailable.

A key limitation of many ML modeling approaches is that standard regression and non-spatial models assume predictor–outcome relationships are stable across geographic space, an assumption known as spatial stationarity. In practice, these relationships vary geographically due to regional differences in population composition, social norms, environmental exposures, healthcare access, and policy contexts, meaning global surrogate models may be too generalized to reliably capture local health dynamics across heterogeneous U.S. counties (Tzavidis et al., 2025). Geographically weighted ML models offer a solution by allowing predictor–outcome relationships to vary across space, and we hypothesize this flexibility makes them better suited as surrogate models for public health SAE.

Therefore, this study explores geographically weighted random forest (GWRF), geographically weighted regression (GWR), and geographical XGBoost (G-XGBoost) as open-data surrogate models for estimating county-level prevalence of ten chronic health outcomes: COPD, asthma, heart disease, arthritis, cancer, depression, diabetes, high blood pressure, high cholesterol, and stroke. The resulting estimates from these models are benchmarked against global comparators including OLS, random forest (RF), and XGBoost. As a case study, we apply our surrogate modeling framework to CDC PLACES, a gold standard SAE reference dataset used in academic research and local and state health policy making. To our knowledge, this is the first study to evaluate ML as an area-level surrogate for a gold-standard health SAE. This study makes two contributions: first, it evaluates whether geographically weighted approaches offer advantages over global models as open-data surrogates for public health SAE; and second, it introduces a transparent, reproducible framework for rapidly generating SAE when commonly used SAE such as CDC PLACES are delayed or unavailable.

## 2. Methods

### 2.1. Methods Overview

We apply a two-stage hierarchical surrogate modeling framework that learns the relationship between county-level social determinants of health (SDOH) and health SAE. The workflow of this surrogate modeling framework applied to the CDC PLACES case study is illustrated in **Figure 1**. In Level 1, models are trained on SDOH to reproduce CDC PLACES county-level smoking and obesity estimates. In Level 2, models are trained on the same SDOH together with smoking and obesity to reproduce CDC PLACES estimates for ten chronic health outcomes. Both levels are trained on 2021-representative data (T1). The reason for the two-stage approach lies in that smoking and obesity are in themselves SAE and they improve subsequent estimations of chronic health outcomes.

To test whether the trained models generalize over time without retraining, the models at both levels are then held fixed and applied to SDOH from later periods (T2 and T3). At each transfer step, the trained Level 1 model first predicts smoking and obesity from the updated SDOH; these predictions are then combined with the same SDOH and passed to the fixed Level 2 model to predict that year's health outcomes. The resulting 2022 and 2023 estimates are validated against the corresponding CDC PLACES data. This experimental design allows us to assess cross-year temporal robustness to determine whether surrogate models can generate reliable estimates in years when updated survey-based SAE products are unavailable (e.g., 2025 and 2026).

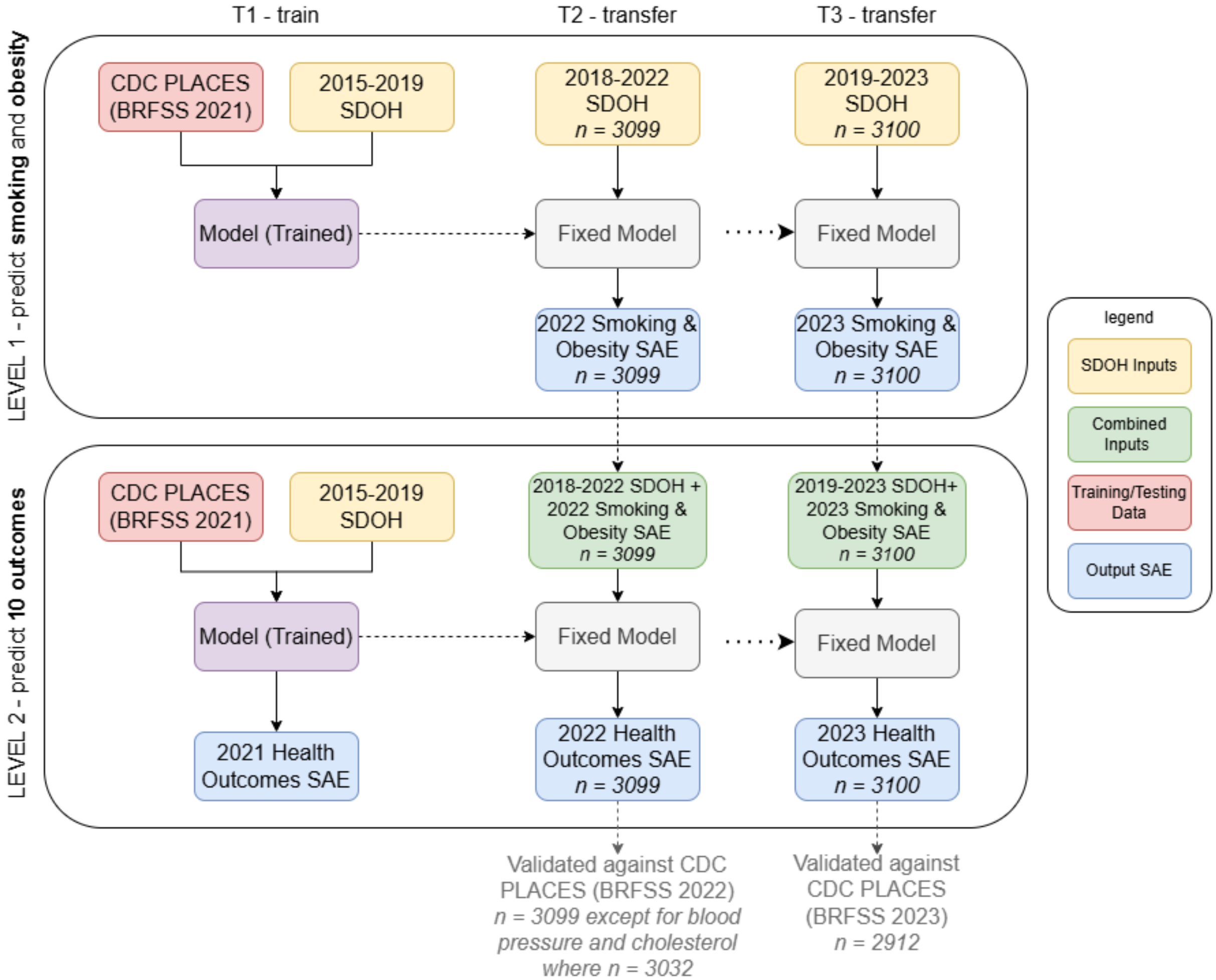

*Figure 1. Two-stage hierarchical ML workflow for SAE of health outcome prevalence across US counties. LEVEL 1 models predict smoking and obesity SAEs from SDOH predictors (age, gender, race/ethnicity, education, poverty, rural population, insurance). LEVEL 2 models combine those SAEs with the same predictors to estimate prevalence of ten outcomes: asthma, heart disease, COPD, stroke, arthritis, cancer, depression, diabetes, high blood pressure, and high cholesterol.*

## 2.2. Data

This study draws on three categories of publicly available, county-level data: CDC PLACES health outcome estimates, ACS- and County Health Rankings (University of Wisconsin Population Health Institute, 2025)–derived SDOH predictors, and U.S. Census Bureau boundary files (U.S. Census Bureau, 2019; U.S. Census Bureau, 2022; U.S. Census Bureau, 2023). For each modeling period (T1, T2, and T3), these sources are aligned to the underlying BRFSS survey year: T1 (2021), T2 (2022), and T3 (2023), rather than the CDC PLACES release year, which has a two-year lag. **Table 1** summarizes the data sources and their vintages across the training and transfer periods.

*Table 1. Summary of input data sources used for model training (T1), temporal transfer testing (T2 and T3), and validation. CDC PLACES provided model-based health outcome estimates with a two year lag, while predictor SDOH variables were derived from ACS 5-year estimates and county health rankings.*

| | Data Source | T1 - train | T2 - transfer | T3 - transfer |
|---|---|---|---|---|
| **Training and Testing Data** | CDC PLACES Release Year | 2023 | 2024 | 2025 |
| | BRFSS | 2021 | 2022 | 2023 |
| **SDOH Predictor Data** | ACS 5 Year Estimates | 2015-2019 | 2018-2022 | 2019-2023 |
| | County Health Rankings | 2019 | 2022 | 2023 |
| **County Boundary Files** | U.S. Census Bureau | 2019 | 2022 | 2023 |

The analysis was restricted to counties with matching FIPS codes in the training and transfer years to maintain a consistent geographic framework across years. Therefore, SAEs were only generated for counties that were present in both T1 and T2 / T3, respectively. Furthermore, we could not generate SAEs for counties that were missing predictor variables in T2 or T3. In total, SAEs were generated for 3099 in T2 and 3100 in T3 out of ~3108 counties (see **Figure 1**). For both T2 and T3, the nine Connecticut planning regions were excluded from the analysis: Capitol, Greater Bridgeport, Lower Connecticut River Valley, Naugatuck Valley, Northeastern Connecticut, Northwest Hills, South Central Connecticut, Southeastern Connecticut, and Western Connecticut. For T2, Oglala Lakota County, South Dakota, was also removed.

### 2.2.1. SDOH Predictor Data

County-level predictor variables were selected to operationalize the five SDOH domains established by Healthy People 2030 (U.S. Department of Health and Human Services, 2020) using measures consistently available from ACS 5-year estimates at the county level. *Economic stability* was captured by percent of the population below the poverty threshold; *educational attainment and quality* by percent of the population with a bachelor's degree or higher; *healthcare access and quality* by percent of uninsured population; *neighborhood and built environment* by percent of the population living in rural areas; and *social and community context* by age composition (percent

≥65), sex (percent female), and racial composition (percent white). Combined, these covariates provide systematic coverage of the social and structural drivers of chronic disease burden at the county level and are consistent with prior SAE and spatial epidemiology studies employing similar covariate frameworks (Rahman, 2017; Viljanen et al., 2022).

Variable selection was further constrained by the need to satisfy the assumptions of the linear models employed in this study, specifically the absence of multicollinearity required by OLS and GWR. Although tree-based models are robust to collinear predictors, a consistent covariate set was maintained across all models to ensure comparability of predictive performance. Variance inflation factors (VIFs) were examined for all candidate variables using the global OLS model, and the final set was chosen to keep VIFs within acceptable thresholds of under 4 (**Supplementary Material, Table S1**). Each variable, its description, and ACS code are provided in **Table 2**.

*Table 2. Categories of SDOH predictor variables derived from the U.S. Census Bureau's ACS 5-Year Estimates and the County Health Rankings and Roadmaps, including the variable names used throughout this paper, the full description, and variable codes in the original data source.*

| SDOH Pillar | Variable Name | Description | Variable Code(s) | Source |
|---|---|---|---|---|
| Social and Community Context | over_65 | Percent of population above the age of 65 | (DP05_0024PE) | American Community Survey from U.S. Census Bureau. (2025) |
| | female | Percent of population that is female | (DP05_0003PE) | |
| | white | Percent of population that identifies as White | (DP05_0037PE) | |
| Education Access and Quality | bach | Percent of population with a bachelor's degree or higher | (S1501_C02_015E) | |
| Economic Stability | poverty | Percent of population with income less than the poverty threshold | (S1701_C03_001E) | |
| Health Care Access and Quality | uninsured | Percent of population that is uninsured | (S2701_C05_001E) | |
| Neighborhood and Built Environment | rural | Percent of the population that is rural | (v058_rawvalue) | County Health Rankings and Roadmaps from University of Wisconsin Population Health Institute. (2025) |

### 2.2.2. Training and Testing Data

To train and test the surrogate models, we used CDC PLACES county-level prevalence for smoking and obesity (LEVEL 1) as well as each of the ten health outcomes (LEVEL 2) among adults aged 18 years or older including COPD, asthma, heart disease, arthritis, cancer, depression, diabetes, high blood pressure, high cholesterol, and stroke (CDC, 2024b). Definitions for each outcome are described on the CDC PLACES website (CDC, 2024c).

## 2.3. Models

### 2.3.1. Global Regression and Machine Learning Models

*Ordinary Least Squares Regression (OLS).* OLS estimates a linear relationship between predictors and outcomes under assumptions of constant error variance and spatial stationarity. It serves as an interpretive and predictive benchmark against which non-linear and spatially varying models are evaluated. The model was implemented using the stats R package (R Core Team, 2024).

*Random Forest (RF).* RF (Breiman, 2001) is an ensemble method that builds decision trees on bootstrap samples of the data, with each split restricted to a random subset of predictors to reduce inter-tree correlation. The model was trained using 800 trees with impurity-based variable importance and mean squared error as the split criterion. The model was implemented using the ranger (Wright and Ziegler, 2017) and caret (Kuhn, 2008) R packages.

*Extreme Gradient Boosting (XGBoost).* XGBoost (Chen & Guestrin, 2016) builds an additive ensemble of trees by iteratively minimizing a differentiable loss function, capturing higher-order feature interactions. Hyperparameters were tuned using five-fold cross-validation. The model was implemented using the xgboost R package (Chen et al., 2025).

### 2.3.2. Geographically Weighted Models

*Geographically Weighted Regression (GWR).* GWR (Brunsdon et al., 1996) extends OLS by allowing regression coefficients to vary spatially, fitting a separate weighted least-squares regression at each location using a spatial kernel that down-weights observations with increasing distance. This relaxes the global stationarity assumption of OLS and captures spatial heterogeneity in the relationships between predictors and outcomes. The model was implemented using the GWmodel R package (Gollini et al., 2015; Lu et al., 2014).

*Geographically Weighted Random Forest (GWRF).* GWRF (Quiñones et al., 2021; Georganos & Kalogirou, 2022) extends the RF framework by integrating spatial weighting, fitting local RF models at each location using a kernel of the 100 nearest neighbors weighted by spatial proximity. This allows predictor-outcome relationships to vary geographically while retaining the non-linear and interaction-capturing properties of RF. The model was implemented using the SpatialML R package (Kalogirou & Georganos, 2024).

*Geographical-XGBoost (G-XGBoost).* G-XGBoost (Grekousis, 2025) extends XGBoost to accommodate spatial non-stationarity through an ensemble architecture combining local and global gradient-boosted models. Distinctively, spatial weights are tied to local prediction error rather than geographic distance alone, concentrating model attention on observations that contribute most to local misfit. The approach also produces location-specific feature importance, enabling spatially varying interpretation of predictor effects. Hyperparameters and spatial bandwidth were tuned through cross-validation. The model was implemented using the geoxgboost Python library (Grekousis, 2025).

### 2.4. Training and Validation

*Training*. For all models except GWR, the T1 county data were split into training (80%) and testing (20%) sets for model construction. GWR was trained using all available counties without a train-test split, allowing the model to utilize the complete spatial dataset.

*Validation.* To evaluate temporal generalizability, surrogate model performance was assessed at T2 and T3 by comparing model-generated SAEs, produced without re-training, to the corresponding CDC PLACES county-level prevalence estimates for each time period. This out-of-sample approach tests the temporal transferability of the learned relationships. For T2, 3099 SAEs were generated. CDC PLACES data were available for validation for all outcomes except high blood pressure and high cholesterol, for which validation data were available for 3032 counties (**Figure 1**). For T3, 3100 SAEs were generated, and CDC PLACES validation data were available for 2912 counties (**Figure 1**).

Two complementary evaluation metrics were used. Pearson's correlation coefficient (r) quantified the strength of the linear association between predicted and CDC PLACES county-level prevalence estimates, where r = 1 indicates perfect agreement. This metric captures how well a model reproduces the spatial distribution of disease burden across counties, or more specifically whether counties with relatively high or low prevalence are ranked correctly, and is the standard evaluation metric used by CDC PLACES for external validation comparisons (Zhang et al., 2015; Wang et al., 2017). Mean Absolute Error (MAE) quantified absolute prediction accuracy in prevalence percentage-point units, with lower values indicating closer agreement with CDC PLACES estimates.

## 3. Results

In this section, we summarize the study's results, beginning with accuracy by health outcome, then comparing global and geographic models, assessing performance across years, and finally benchmarking against direct estimates in a case study. The full, non-aggregated results are provided in **Supplementary Materials S2 and S3.**

### 3.1. Performance by Health Outcome

**Table 3** summarizes the Pearson's $r$ and MAE between the surrogate model estimates and CDC PLACES for each disease outcome, across all models and all years to show which outcomes were easier to estimate. The closest average agreement ($r > 0.9$, MAE < 1%) was for stroke ($r \approx 0.94$, MAE ≈ 0.38), heart disease ($r \approx 0.93$, MAE ≈ 0.60), and COPD ($r \approx 0.90$, MAE ≈ 0.78). A second set of outcomes still aligned strongly ($r > 0.9$) but with MAE under 2%, including as follows: diabetes ($r \approx 0.91$, MAE ≈ 1.05), cancer ($r \approx 0.92$, MAE ≈ 1.13), and high blood pressure ($r \approx 0.90$, MAE ≈ 1.82). High cholesterol ($r \approx 0.82$, MAE ≈1.62) and arthritis ($r \approx 0.88$, MAE ≈ 1.85) showed moderate agreement, with lower correlations and mid-range error. Depression and asthma had the weakest correlations ($r \approx 0.70$ and 0.67). Depression also had the highest error (MAE ≈ 2.27), but asthma was very low (MAE ≈ 0.69).

*Table 3. Summary table of model performance metrics for predicting county-level health outcome prevalence across the US using T2 and T3 datasets, summarized for each health outcome.*

| Health Outcome | Pearson's r | | | | MAE | | | |
|---|---|---|---|---|---|---|---|---|
| | Average | Standard Deviation | Minimum | Maximum | Average | Standard Deviation | Minimum | Maximum |
| Arthritis | 0.878 | 0.034 | 0.814 | 0.921 | 1.845 | 0.402 | 1.293 | 2.474 |
| Asthma | 0.665 | 0.122 | 0.446 | 0.812 | 0.685 | 0.062 | 0.605 | 0.778 |
| Cancer | 0.924 | 0.010 | 0.909 | 0.938 | 1.130 | 0.054 | 1.065 | 1.211 |
| COPD | 0.901 | 0.035 | 0.821 | 0.943 | 0.779 | 0.210 | 0.512 | 1.165 |
| Depression | 0.702 | 0.121 | 0.527 | 0.865 | 2.270 | 0.395 | 1.649 | 2.864 |
| Diabetes | 0.906 | 0.026 | 0.854 | 0.942 | 1.045 | 0.106 | 0.834 | 1.191 |
| Heart Disease | 0.927 | 0.018 | 0.900 | 0.950 | 0.604 | 0.229 | 0.332 | 0.860 |
| High Blood Pressure | 0.901 | 0.048 | 0.825 | 0.960 | 1.816 | 0.435 | 1.160 | 2.405 |
| High Cholesterol | 0.815 | 0.066 | 0.706 | 0.929 | 1.617 | 0.176 | 1.360 | 1.924 |
| Stroke | 0.935 | 0.015 | 0.908 | 0.957 | 0.380 | 0.089 | 0.251 | 0.489 |

### 3.2. Model Comparison: Global Models vs. Geographic Models

**Table 4** reports the same summarized metrics for each model across all outcomes and years to compare performance across the different employed global and geographic models. Models with a higher average $r$ also tended to have a lower average MAE. Among the global models, XGBoost performed best ($r \approx 0.84$, MAE = 1.30), ahead of RF ($r \approx 0.83$, MAE ≈ 1.31) and OLS ($r \approx 0.79$, MAE ≈ 1.42). Among the geographic models, GWR and GWRF performed similarly ($r \approx 0.90$ and 0.90; MAE ≈ 1.07 and 1.09), followed by G-XGBoost ($r \approx 0.88$, MAE = 1.11). Overall, the models that allow for spatial non-stationarity outperformed their global counterparts on average.

*Table 4. Summary table of model performance metrics for predicting county-level health outcome prevalence across the US using T2 and T3 datasets, summarized for each type of model.*

| Model | Pearson's r | | | | MAE | | | |
|---|---|---|---|---|---|---|---|---|
| | Average | Standard Deviation | Minimum | Maximum | Average | Standard Deviation | Minimum | Maximum |
| OLS | 0.788 | 0.157 | 0.446 | 0.940 | 1.420 | 0.829 | 0.285 | 2.864 |
| GWR | 0.899 | 0.059 | 0.757 | 0.957 | 1.073 | 0.563 | 0.251 | 2.196 |
| RF | 0.830 | 0.125 | 0.545 | 0.947 | 1.314 | 0.700 | 0.325 | 2.491 |
| GWRF | 0.897 | 0.050 | 0.781 | 0.960 | 1.090 | 0.500 | 0.341 | 2.042 |
| XGBoost | 0.836 | 0.122 | 0.575 | 0.952 | 1.296 | 0.712 | 0.288 | 2.545 |
| G-XGBoost | 0.882 | 0.073 | 0.720 | 0.954 | 1.109 | 0.548 | 0.302 | 2.025 |

Disaggregating model performance by disease outcome reveals that the advantage for using geographical models is highly uneven. Comparing each global model to its geographic counterpart across both transfer years (**Figure 2**), the gain in average Pearson's *r* is largest for depression (global ≈ 0.59, geographic ≈ 0.81; Δ ≈ 0.22) and asthma (≈ 0.56 vs. ≈ 0.77; Δ ≈ 0.21), followed by high cholesterol (≈ 0.77 vs. ≈ 0.87; Δ ≈ 0.10) and high blood pressure (≈ 0.86 vs. ≈ 0.94; Δ ≈ 0.08). The largest reductions in MAE occur for high blood pressure (global ≈ 2.19, geographic ≈ 1.44; Δ ≈ −0.74) and depression (global ≈ 2.61, geographic ≈ 1.93; Δ ≈ −0.67).

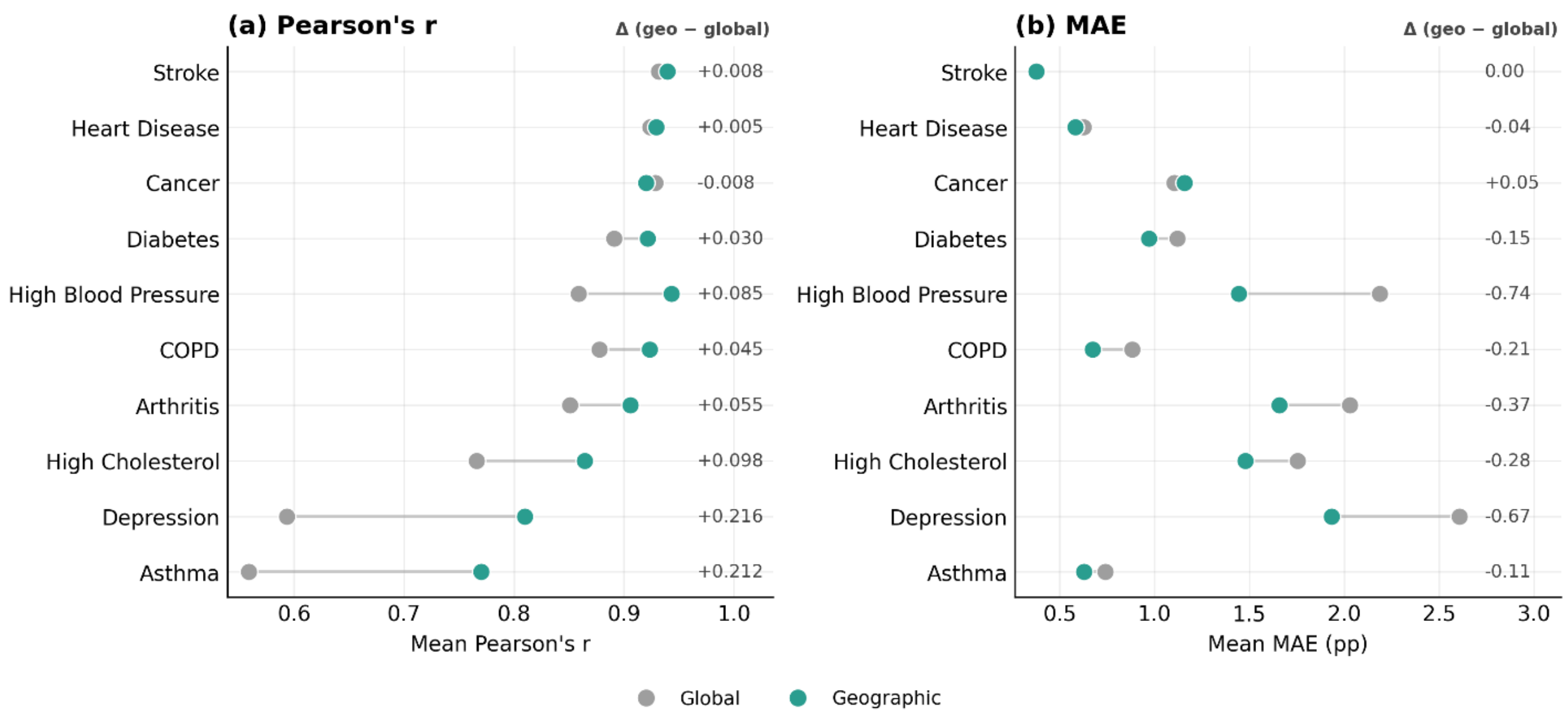

*Figure 2. Global versus geographic model performance by disease outcome. For each outcome, points show the mean Pearson's r (a) and mean MAE percentage points (pp) (b) of the three global models (OLS, random forest, XGBoost; grey) and their three geographic counterparts (GWR, GWRF, G-XGBoost; green), averaged across both transfer years (T2 and T3).*

The clearest illustration of this advantage is depression, the outcome with the largest average gain in performance between the global models and the geographic models. Because global models assume a single predictor–outcome relationship everywhere, they cannot capture regional differences in how social determinants relate to depression prevalence, and this failure leaves a spatial signature in their errors. **Figure 3** maps county-level residuals for depression at T2 for a global model with the worst performance (OLS, **Figure 3a**) and a geographic model with the best performance (GWRF, **Figure 3b**), where positive residuals (red) indicate underestimation and negative residuals (blue) overestimation. Under OLS, the residuals are highly spatially clustered indicating that a single global relationship does not fit all parts of the country. GWRF, by allowing the predictor–outcome relationship to vary across space, substantially reduces this clustering, producing residuals that are smaller and more spatially random. This is the mechanism behind depression's large performance gain: the spatial structure that global models leave in their errors is precisely what the geographically weighted models absorb.

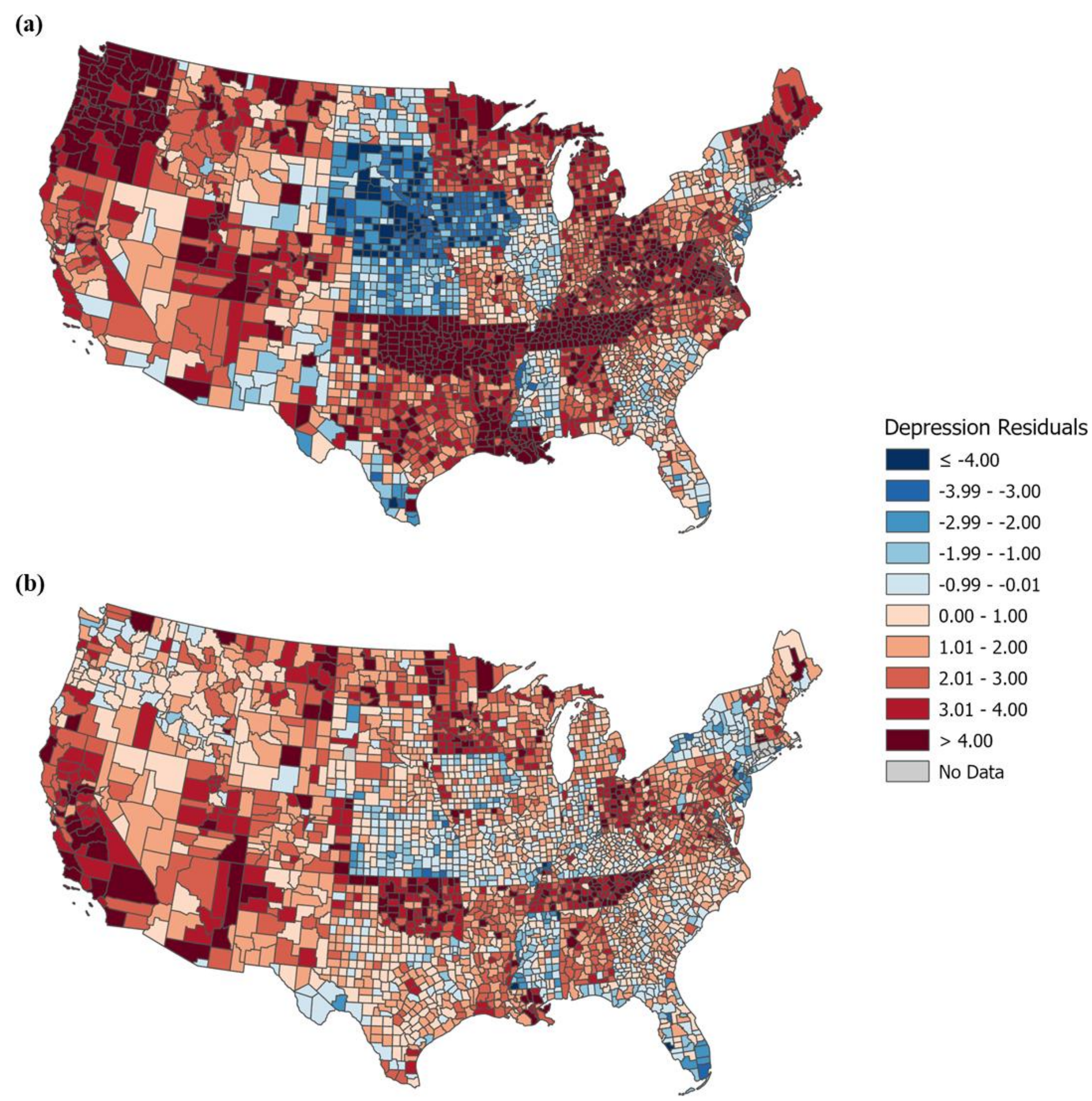

*Figure 3. Map of U.S. county surrogate model residuals for depression using OLS (a) and GWRF (b) for T2. Map was created using ArcGIS Pro 3.6.2. (Esri, 2026) and openly available boundary files from the U.S. Census Bureau (U.S. Census Bureau, 2022).*

For outcomes that global models already estimate well, the geographic models do not perform much better: stroke and heart disease are essentially unchanged, and the performance of the geographic models marginally decrease than their global counterparts for cancer ($\Delta r \approx -0.01$, $\Delta$MAE $\approx 0.05$). Because the largest gains fall on the outcomes that the global models estimate worst, the geographic models also narrow the range of performance across outcomes, from roughly 0.56–0.93 down to 0.77–0.94 for Pearson's $r$ correlation coefficient and from 0.38-2.6 down to 0.38-1.93 for MAE. In practical terms, geographic model accuracy becomes far less dependent on which outcome is estimated.

**Figure 4** shows the performance for all models across health outcomes. Of the global models, XGBoost performed consistently better, with differences in performance across models widening for outcomes that were harder to estimate (**Figure 4, grey**); however, the best performing geographic models ultimately depended on the outcome. Across the geographic models (**Figure 4, green**), GWR (green circle) produced the highest values for the outcomes with strong overall agreement, including cancer ($r \approx 0.93$, MAE ≈ 1.08), stroke ($r \approx 0.94$, MAE ≈ 0.34), heart disease ($r \approx 0.94$, MAE ≈ 0.54), and diabetes ($r \approx 0.93$, MAE ≈ 0.87). For the more difficult outcomes, however, GWRF (green square) had a Pearson's $r$ that exceeded GWR for depression (≈ 0.84 vs. ≈ 0.82), asthma (≈ 0.80 vs. ≈ 0.78), and high cholesterol (≈ 0.88 vs. ≈ 0.87), although in general GWR maintained the lowest MAE. G-XGBoost (green triangle) was the lowest-performing geographic model for most outcomes, and this gap was widest for the hardest outcomes, where its Pearson's $r$ values fell below those of both GWR and GWRF for depression ($r \approx 0.78$, MAE ≈ 1.96) and asthma ($r \approx 0.74$, MAE ≈ 0.62), although its MAE was intermediate rather than the worst.

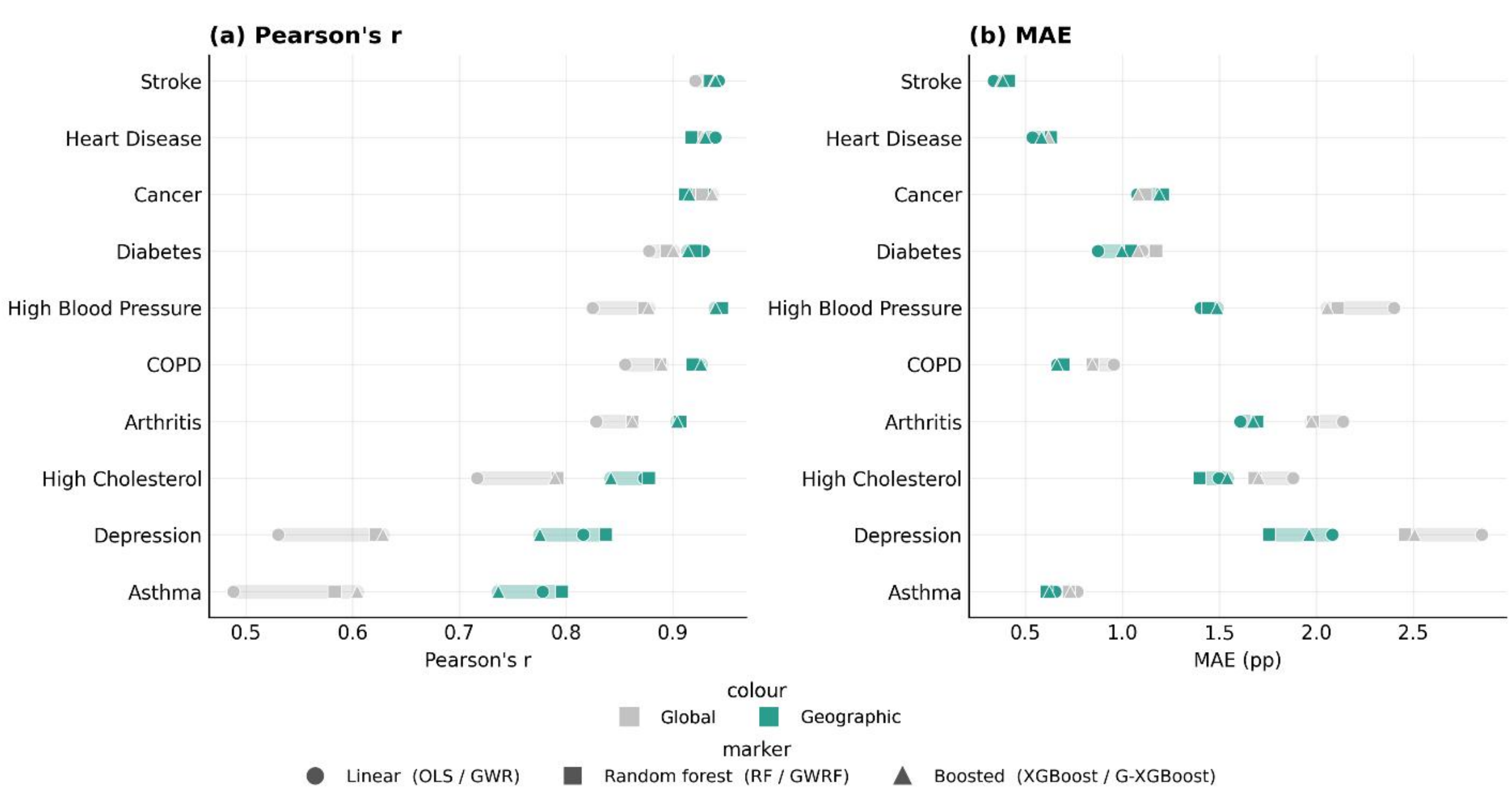

*Figure 4. Per-outcome Pearson's r (a) and MAE percentage points (pp) (b) for all six models, averaged over transfer years T2 and T3. Color distinguishes global from geographic models; marker shape distinguishes the underlying model. The shaded bar shows the range across the three models within each family.*

### 3.3. Transferability Across Years

Across all models, differences between the two transfer years, T2 and T3, were small for the well-estimated outcomes and largest for the outcomes that were more difficult to estimate accurately (**Figure 5**). The best-estimated outcomes (e.g., stroke, heart disease, diabetes, COPD, and arthritis) were stable or improved slightly, with higher Pearson's *r* and lower MAE, whereas the outcomes that are more difficult to estimate (e.g., high cholesterol, high blood pressure, depression, and asthma) declined between the two years, with lower Pearson's *r* and stable to moderate increases in MAE. GWRF (green square) showed the largest year-to-year variation of the geographic models, with its high cholesterol Pearson's *r* dropping by more than 0.10 and its high blood pressure MAE increasing slightly between transfer years.

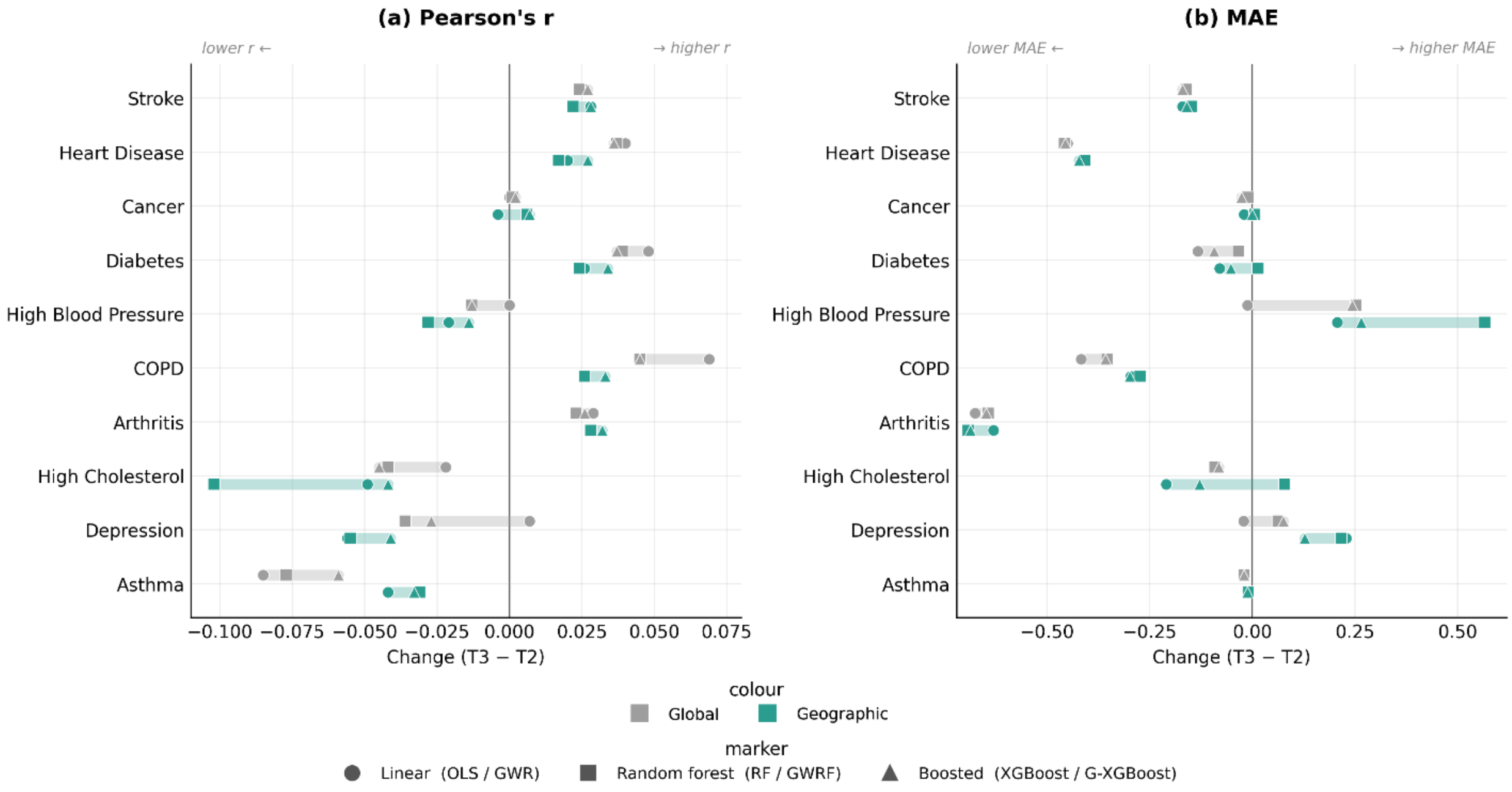

*Figure 5. Change in model performance between the two transfer years for Pearson's r (a) and MAE (b). For each outcome, points show the change from T2 to T3 (T3 − T2) for each model. Within each outcome, the three global models (grey, upper row) and three geographic models (green, lower row) are shown separately, with the shaded bar spanning the range across the models in each family. Marker shape denotes the algorithm (circle = linear, square = random forest, triangle = boosted).*

### 3.4. Case Study with Direct Estimates

As a final validation, we compared both the surrogate model estimates and the CDC PLACES estimates against direct BRFSS survey estimates at the county level in Arkansas for T2. We use Arkansas because it was the only state we could find direct estimates for our transfer years. We selected three outcomes, COPD, asthma, and heart disease, chosen to span the range of agreement with CDC PLACES observed in the primary analysis, from among the best-estimated outcomes to the weakest (**Table 5**).

For COPD, agreement with the direct estimates was comparable between our surrogate model and CDC PLACES. The surrogate models ranged from $r = 0.559$ to 0.637, compared to CDC PLACES at $r = 0.611$; the highest agreement with direct estimates came from global XGBoost. MAE values were also similar, ranging from 1.591 to 1.793 for the surrogate models compared to 1.611 for CDC PLACES. For heart disease, the surrogate models ranged from $r = 0.394$ to 0.475 against the direct estimates, with the strongest model (G-XGBoost) slightly exceeding CDC PLACES ($r = 0.425$). MAE values favored the surrogate models, ranging from 1.185 to 1.555 compared to 2.397 for CDC PLACES. For asthma, neither the surrogate model nor CDC PLACES showed meaningful agreement with the direct estimates. The surrogate models produced Pearson's $r$ values ranging from −0.116 to −0.004, and CDC PLACES produced a value of 0.010. However, surrogate models had slightly lower MAE values, ranging from 1.737 to 1.825 compared to 1.951 for CDC PLACES. Across the three outcomes, no surrogate model type consistently outperformed the others against the direct estimates.

*Table 5. Pearson's r and MAE values for county-level health outcome prevalence estimates in Arkansas for T2: surrogate model estimates vs. BRFSS estimates and CDC PLACES estimates vs. BRFSS estimates.*

| **Model** | | **Arkansas T2 Estimates** | | | |
|---|---|---|---|---|---|
| | | **Pearson's r** | | **MAE** | |
| | | **Surrogate Models vs. BRFSS** | **CDC PLACES vs. BRFSS** | **Surrogate Models vs. BRFSS** | **CDC PLACES vs. BRFSS** |
| **COPD** | OLS | 0.594 | 0.611 | 1.707 | 1.611 |
| | GWR | 0.597 | | 1.631 | |
| | RF | 0.559 | | 1.793 | |
| | GWRF | 0.612 | | 1.591 | |
| | XGBoost | 0.637 | | 1.620 | |
| | G-XGBoost | 0.611 | | 1.610 | |
| **Asthma** | OLS | -0.106 | 0.010 | 1.825 | 1.951 |
| | GWR | -0.094 | | 1.757 | |
| | RF | -0.059 | | 1.771 | |
| | GWRF | -0.116 | | 1.737 | |
| | XGBoost | -0.061 | | 1.790 | |
| | G-XGBoost | -0.004 | | 1.744 | |
| **Heart Disease** | OLS | 0.465 | 0.425 | 1.185 | 2.397 |
| | GWR | 0.468 | | 1.533 | |
| | RF | 0.444 | | 1.245 | |
| | GWRF | 0.394 | | 1.555 | |
| | XGBoost | 0.447 | | 1.275 | |
| | G-XGBoost | 0.475 | | 1.444 | |

**Figure 6** compares county-level BRFSS heart disease direct estimates (**Figure 6a**) with CDC PLACES estimates (**Figure 6b**) and surrogate model estimates from G-XGBoost (**Figure 6c**) for T2. Relative to the BRFSS direct estimates, both CDC PLACES and G-XGBoost reproduce the broad spatial pattern of heart disease prevalence in Arkansas, including higher prevalence in northeastern and southwestern counties and lower prevalence in parts of central Arkansas. However, compared with BRFSS, CDC PLACES produces a set of estimates that shifts many counties into the moderate to high range and smooths out local extremes. G-XGBoost produces more of the county-level variability and overall prevalence level of the direct estimates, increasing Pearson's r and reducing MAE compared to CDC PLACES.

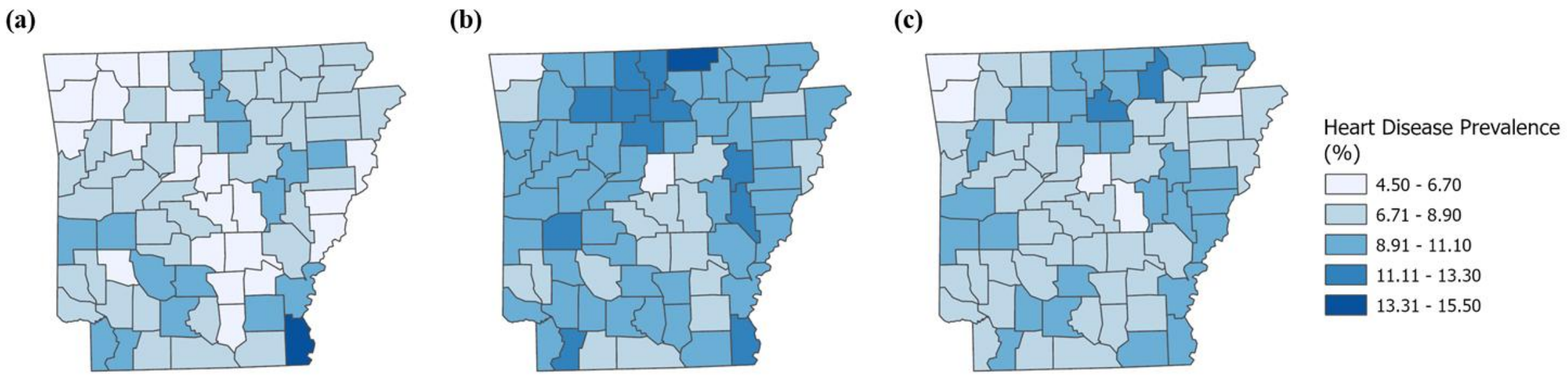

*Figure 6. Map of Arkansas heart disease from BRFSS estimates (a), CDC PLACES estimates (b), and surrogate model estimates (c) using G-XGBoost for T2. Map was created using ArcGIS Pro 3.6.2. (Esri, 2026) and openly available boundary files from the U.S. Census Bureau (U.S. Census Bureau, 2022).*

## 4. Discussion and Conclusion

This study shows that we can offer a transparent, open-data surrogate that can stand in for SAE in the years when survey releases create a lag. Using publicly available county-level predictors, the surrogate models produced CDC PLACES-like estimates for chronic disease outcomes and transferred successfully across two subsequent years without retraining. GWR performed best on average, with the highest overall Pearson's correlation and lowest MAE. This may have something to do with the fact that CDC PLACES uses a linear model, meaning that a local linear surrogate may be well suited to reproducing its estimates. If the target estimates are partly shaped by that structure, then a local linear model such as GWR may be enough to recover much of the relationship. However, GWRF nearly matched GWR for the easier outcomes and performed better for several outcomes that were more difficult to predict. This suggests that GWRF may be better suited for applications that require estimating multiple outcomes because its results are less sensitive to the specific outcome being modeled. This makes the estimates more reliable for public health applications because performance is less tied to the specific disease being modeled.

Our results show why spatial non-stationarity matters. The improvement from geographic models was large for depression, asthma, high cholesterol, and high blood pressure, while stroke, heart disease, and cancer were almost unchanged between global and geographic models. This suggests that outcomes whose predictor-outcome relationships are demographically driven and roughly stable across space are already well captured by global models, so locality provides limited additional benefit. Outcomes with more contingent drivers, environmental exposures, healthcare access, and regional norms, have spatially varying relationships that global models may miss. These spatially varying relationships may be captured as a latent component in the geographic models.

The temporal transfer results suggest that surrogate models can remain reliable across years, but not equally for all outcomes. Spatial models stayed stable or improved on the well-estimated outcomes across T2 to T3, while harder outcomes declined. GWRF showed the largest year-to-year swing, which is important when considering it as a practical recommendation. Although GWRF may be less sensitive to the outcome being modeled, it may be more sensitive to temporal instability.

The Arkansas case study showed that the surrogate estimates generally agreed with direct BRFSS estimates at a level comparable to CDC PLACES. However, this agreement varied by outcome. For asthma, neither product showed meaningful agreement with the direct estimates. This suggests that the surrogate approach may be reaching a similar accuracy ceiling as CDC PLACES, and it may also face similar limits in cases where the available data do not capture the outcome well.

These findings have practical and policy-relevant implications. The surrogate framework is timely, open, reproducible, scalable to U.S. counties, and does not require individual-level microdata. It can provide a bridge for health departments in years when CDC PLACES SAE is not yet available. The results also provide evidence that SAE pipelines should incorporate spatial non-stationarity, especially for outcomes where conventional global models fail.

At the same time, several limitations should be acknowledged. Because the surrogate models are trained on PLACES data, they inherit PLACES biases. Additionally, this study only explores performance of ML models for generating county-level estimates using a small set of predictor variables. There is potential to support estimation at tract or block-group scales and expand the set of predictor variables, something that will be explored in future work. Furthermore, the models in this study were tested across only two transfer years, and the geographical modeling approaches can only account for spatial but not temporal non-stationarity in the relationships themselves. Finally, GWRF's greater temporal instability is also a limitation. Although GWRF performed well for several harder outcomes, its larger year-to-year changes suggest that its transferability may be less stable than the other models.

Future work should focus on testing how well this surrogate framework generalizes beyond the current study design. This includes evaluating richer predictors, extending the approach to finer geographies, and modeling temporal drift over longer periods. A natural extension of this work is to apply the surrogate modeling framework to SAE products and estimate types beyond CDC PLACES. Direct survey estimates present a particularly compelling case: available only for a subset of states and only intermittently, they could in principle be reproduced by a surrogate model to bridge the resulting temporal and geographic gaps. Such an application, however, confronts a fundamental evaluation problem, namely the absence of a reliable ground truth. CDC PLACES is itself validated against direct estimates, yet these are subject to considerable sampling noise, as our Arkansas case study illustrates, rendering them an imperfect benchmark. Any surrogate targeting direct estimates must therefore contend with the question of what constitutes acceptable agreement when the validation target itself is uncertain. The framework could likewise be extended to other estimates released only intermittently, such as county-level or finer-scale influenza vaccination coverage, to test its generality across public health surveillance products.

In conclusion, spatially weighted surrogate models offer accurate, timely, open-data county-level chronic disease estimates that remain reliable across years. Their value is greatest where conventional global models performance is weaker, suggesting that spatial non-stationarity matters most for outcomes shaped by spatially varying relationships. These models should not be understood as replacements for survey-based SAE, but as a way to generate timely estimates when products such as CDC PLACES are delayed or unavailable. Overall, the results show that geographically weighted surrogate models can support public health surveillance while also providing evidence that future SAE pipelines should more directly account for spatially varying predictor-outcome relationships.